\documentclass[twocolumn]{aastex63}

\usepackage[english]{babel}
\usepackage[utf8x]{inputenc}
\usepackage[T1]{fontenc}

\usepackage{amsmath}
\usepackage{graphicx}
\usepackage{upgreek}
\usepackage{ulem}
\newcommand{\radu}{rad m$^{-2}$}

\shorttitle{CGPS Rotation measure Catalogue}
\shortauthors{Van Eck et al. }

\begin{document}

\title{Revisiting Rotation Measures from the Canadian Galactic Plane Survey: the Magnetic Field in the Disk of the Outer Galaxy}
\correspondingauthor{C.L. Van Eck}
\email{cameron.van.eck@dunlap.utoronto.ca}

\author[0000-0002-7641-9946]{C.L. Van Eck}
\affiliation{Dunlap Institute for Astronomy and Astrophysics, University of Toronto, 50 St. George Street, Toronto, ON M5S 3H4, Canada}

\author[0000-0003-4781-5701]{J.C. Brown}
\affiliation{Department of Physics and Astronomy, University of Calgary,
    Calgary, Alberta, Canada T2N 1N4}
    
\author[0000-0002-2465-8937]{A. Ordog}
\affiliation{Department of Physics and Astronomy, University of Calgary,
    Calgary, Alberta, Canada T2N 1N4}
\affiliation{Department of Computer Science, Math, Physics, \& Statistics, Irving K. Barber Faculty of Science, University of British Columbia, Okanagan Campus, Kelowna, BC V1V 1V7 Canada}
\affiliation{National Research Council Canada, Herzberg Research Centre for Astronomy and Astrophysics, Dominion Radio Astrophysical Observatory, PO Box 248, Penticton, BC, Canada V2A 6J9}

\author[0000-0001-5953-0100]{R. Kothes}
\author[0000-0003-1455-2546]{T.L. Landecker}
\affiliation{National Research Council Canada, Herzberg Research Centre for Astronomy and Astrophysics, Dominion Radio Astrophysical Observatory, PO Box 248, Penticton, BC, Canada V2A 6J9}

\author{B. Cooper}
\author{K.M. Rae}
\affiliation{Department of Physics and Astronomy, University of Calgary,
    Calgary, Alberta, Canada T2N 1N4}
  
\author{D.A. Del Rizzo}
\author[0000-0002-2280-7644]{A.D. Gray}
\affiliation{National Research Council Canada, Herzberg Research Centre for Astronomy and Astrophysics, Dominion Radio Astrophysical Observatory, PO Box 248, Penticton, BC, Canada V2A 6J9}
\author[0000-0003-2469-1611]{R. Ransom}
\affiliation{Department of Physics and Astronomy, Okanagan College, 583 Duncan Avenue West, Penticton, BC V2A 8E1, Canada}
\affiliation{National Research Council Canada, Herzberg Research Centre for Astronomy and Astrophysics, Dominion Radio Astrophysical Observatory, PO Box 248, Penticton, BC, Canada V2A 6J9}
\author[0000-0003-2391-8650]{R.I~Reid}
\affiliation{Department of Information Technology, Mayo Clinic and Foundation, 200 1st Street SW, Rochester, MN, USA 55905}
\author{B. Uyaniker}
\affiliation{DataSpeckle Scientific Inc., Kelowna, BC, Canada}

\begin{abstract}
Faraday rotation provides a valuable tracer of magnetic fields in the interstellar medium; catalogs of Faraday rotation measures provide key observations for studies of the Galactic magnetic field. We present a new catalog of rotation measures derived from the Canadian Galactic Plane Survey, covering a large region of the Galactic plane spanning 52\degr < $l$ < 192\degr, -3\degr\ < $b$ < 5\degr, along with   northern and southern latitude extensions around $l\approx$ 105\degr. We have derived rotation measures for 2234 sources (4 of which are known pulsars), 75\% of which have no previous measurements, over an area of approximately 1300 square degrees. These new rotation measures increase the measurement density for this region of the Galactic plane by a factor of two.
\end{abstract}

\section{Introduction}
Magnetic fields are an important component of the interstellar medium (ISM) in terms of dynamics and evolution, with typical energy densities comparable to the turbulence and significantly exceeding the thermal energy \citep{Beck2007}. As a result, magnetic fields have been found to play a role in many ISM processes, including the acceleration and confinement of cosmic rays \citep{Aharonian2012}, cloud collapse in the early stages of star formation \citep{Padoan2011}, and the vertical structure of the Galactic disk \citep{Boulares90}.

Several different methods of observing magnetic fields in the ISM have been used to explore different properties of these magnetic fields. Among these is the measurement of Faraday rotation, which has been used to study large scale structure in Galactic magnetic fields \citep[e.g.,][]{Han2006,Brown07,Sun2008}. Faraday rotation measures (RMs) of polarized radio sources give the strength of the Faraday rotation between a source and the observer, which in turn gives information on the magnetic field along the line of sight. RMs are determined by observing how the polarization angle changes with frequency; for a Faraday-simple source\footnote{Faraday-simple refers to a line of sight with a single dominant source of polarized emission, with all of that emission undergoing the same amount of Faraday rotation. Lines of sight with multiple polarized components with different Faraday rotation are called Faraday-complex.}, the relationship between polarization angle ($\psi$) and wavelength ($\lambda$) is
\begin{eqnarray*}
\psi &=& \psi_0 + \lambda^2 \left[ 0.812\; \mathrm{rad\; m}^{-2} \int_{\vec{d}}^{0} \left( \frac{n_\mathrm{e}}{{\rm cm^{-3}}}\right) \left( \frac{\vec{B}}{{\rm \upmu G}} \right) \cdot \left(\frac{\vec{dl}}{{\rm pc}} \right) \right] \\
& = & \psi_0 + \lambda^2 \; \mathrm{RM},
\end{eqnarray*}
where $\psi_0$ is the polarization angle before Faraday rotation, $n_e$ is the density of free electrons, $\vec{B}$ is the magnetic field, $\vec{dl}$ is a differential element of the radiation path, and the integral is evaluated along the path from the source at a location $\vec{d}$ to the observer.

RMs of pulsars and extragalactic sources thus provide information on the magnetic field and free electron density along the line of sight to each source. Combined with models of the free electron density (determined through other tracers such as dispersion measures), RMs can be used to construct models of the magnetic field. Since each RM gives information for only its specific line of sight, and includes contributions from the small-scale fluctuations in the magnetic field and free electron density, many RMs on adjacent lines of sight (and at different distances, when pulsars are used) are needed to disentangle structures on different scales and to statistically remove the effects of small scale fluctuations. For this reason having a high sky density of RMs is very important in maximizing the value and reliability of analysis using RMs, and surveys that produce large numbers of RMs are needed.

Faraday rotation measure surveys can be loosely divided into three generations. The first generation RM surveys, encompassing most of those surveys conducted prior to the 1990's, are characterized by small numbers of objects (typically $\log(N)\approx 1-2$), very few frequency channels (often observed asynchronously and with different receivers) with a correspondingly high vulnerability to $n\pi$ ambiguities, and low sky-surface density of RMs. Examples of such surveys are \citet{Vallee75}, \citet{Rudnick1983}, and \citet{Clegg1992}. The second generation  RM surveys improved on this by having larger source counts ($\log(N)\approx 2-3$), greater bandwidth divided into more frequency channels to overcome $n \pi$ ambiguities and in some cases to enable the use of RM synthesis \citep{Brentjens05}, as well as higher sky density and/or larger sky coverage. Examples include \citet{Brown2003}, \citet{Brown07}, \citet{Taylor09}, \citet{Mao2010}, and \citet{VanEck11}.

The third generation RM surveys, which have just begun, take full advantage of advances in receiver and computer technology to have hundreds to thousands of frequency channels with high total fractional bandwidth, allowing the use of RM synthesis and other advanced techniques to explore additional polarization properties like Faraday complexity \citep[e.g.,][]{Brown2018}. This generation will include large all-sky surveys, in particular the Very Large Array Sky Survey \citep[VLASS,][]{Mao2014}, the POlarization Sky Survey of the Universe's Magnetism \citep[POSSUM,][]{Gaensler2010}, and ultimately surveys with the Square Kilometre Array \citep[SKA,][]{Heald20}. Each of those surveys will massively increase the number of measured RMs ($\log(N)\approx 5-6.5$). The third generation will also include high-precision RM surveys using low frequency radio telescopes \citep[e.g.,][]{VanEck2018a,Riseley2018}, and surveys with very high fractional bandwidths \citep[e.g.,][]{Shanahan2019, Schnitzeler2019, Ma2020}.

The Canadian Galactic Plane Survey (CGPS) was a multi-wavelength observing campaign to study the interstellar medium in the region of the Galactic plane visible from the northern hemisphere \citep{Taylor2003}. This included a radio polarization survey performed using the Synthesis Telescope at the Dominion Radio Astrophysical Observatory \citep[DRAO-ST;][]{Landecker2010}. Observations for the survey began in 1995 and continued in several phases until 2009, and covered the Galactic disk from 52\degr\ < $l$ < 192\degr, -3\degr\ < $b$ < 5\degr\ as well as an extension to higher latitudes above the plane covering 101\degr\ < $l$ < 116\degr, 5\degr\ < $b$ < 17.5\degr. Later observations added a southern latitude extension (SLE) covering 100\degr\ < $l$ < 111\degr, -10\degr\ < $b$ < -3\degr. A rotation measure catalog of part of the first phase of the CGPS (82\degr\ < $l$ < 96\degr, 115\degr\ < $l$ < 147\degr) was produced by \citet{Brown2003}.

In this paper we report on the search for compact polarized sources with reliable RMs within the full CGPS and SLE data, and present a catalog of the resulting measurements; this may be the last large catalog of the second generation RM surveys. In Sect.~\ref{sec:data} we describe the CGPS polarization data in more detail, and describe our method of determining which sources had significant polarization and measuring their RMs. In Sect.~\ref{sec:catalog} we present the resulting catalog of 2234 polarized sources, and compare this catalog with previously reported RM data. Sect.~\ref{sec:analysis} presents some analysis of the Galactic magnetic field using our catalog and demonstrates the unique value of this catalog. Finally we summarize and present our conclusions in Sect.~\ref{sec:summary}.

\section{Data and RM determination}\label{sec:data}
\subsection{CGPS data}
The details of the CGPS 1.4 GHz observations and subsequent processing through to the final images are reported in full detail by \citet{Landecker2010}; we highlight a few key parameters in Table~\ref{table:data}. For our analysis, we have used the data from the DRAO-ST only, without the single-dish data from the DRAO 26-m John Galt telescope and Effelsberg 100-m telescope. The data for the SLE region were processed in the same way as the CGPS data. The data were supplied in the form of mosaic images, each 5.1\degr\ square, with a single channel-averaged Stokes $I$ image and Stokes $Q$ and $U$ images for each channel. The SLE region was combined into a single 10\degr\ square mosaic.

\begin{table}\label{table:data}
    \centering
    \begin{tabular}{|c|c|} \hline
      {\bf Parameter}   & {\bf Value} \\ \hline
        \# of channels & 4\\
        Channel Bandwidth & 7.5 MHz\\
        Center frequencies & 1407.2, 1414.1, \\
                         & 1427.7, and 1434.6 MHz\\
        Beam size & 58\arcsec $\times$ 58\arcsec\ cosec $\delta$\\
        Nominal sensitivity & 0.23-0.30 mJy/beam rms\\ \hline
    \end{tabular}
    \caption{Key Observational Parameters for the CGPS Data.}
    
\end{table}

\subsection{Polarized source identification}

\begin{figure*}[p]
    \centering
    \includegraphics[width=\linewidth,height=0.9\textheight,keepaspectratio]{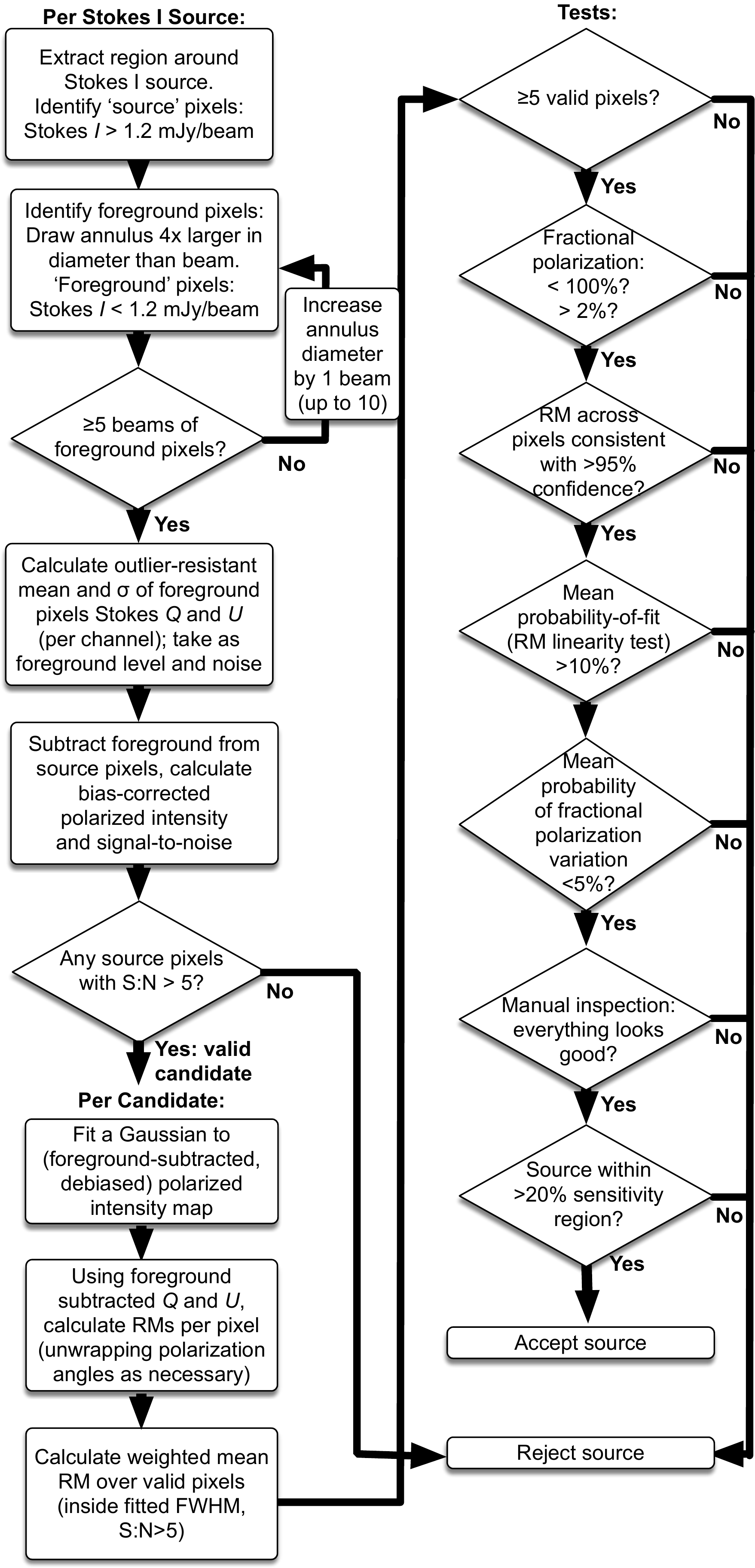}
    \caption{A flowchart showing the key steps in identifying polarized source candidates and evaluating the reliability of their calculated rotation measure. Details of each step are given in the text.}
    \label{fig:flowchart}
\end{figure*}

The identification of polarized sources and determination of their RMs followed the method of \citet{Brown2003}, with some improvements. For completeness, the method is described in full below, and a flowchart of the method is shown in Fig.~\ref{fig:flowchart}.

The first step was the identification of polarized source candidates. We began with a Stokes $I$ source list, and searching each source location for statistically significant polarized intensity. Within the main CGPS region, we used the CGPS Stokes $I$ source catalog from \citet{Taylor2017} as the input source list; for the SLE region we used the {\it Aegean} source-finder \citep{Hancock2012,Hancock2018} with the default parameters to generate an input source list.

Each Stokes $I$ source in the input lists was assessed for statistically significant polarized emission on the source location. This was made difficult by the combination of two related challenges: the presence of diffuse polarized emission in many regions, as well as the position-dependent noise level across the survey. To account for these factors, we determined the local noise and foreground around each source. The local off-source region around each source was defined as an ellipse with the shape of the synthesized beam (calculated at the source location, as the beam shape changes significantly with declination) and size equal to 4 times the full width at half maximum (FWHM) of the beam in each dimension (producing a beam-shaped region with an area of 16 beams). Pixels within this region with Stokes $I$ levels above 1.2 mJy/beam (approximately 5 sigma across most of the survey) were classified as `source pixels' and those below this threshold as `off-source/foreground pixels'; this prevented any polarization from the target source or neighbouring sources from being included in the foreground calculations. Before calculating the noise and foreground in the off-source pixels, we required that there be at least 5 beam-areas worth of pixels in the off-source region; if there were insufficient pixels we increased the size of the local-region ellipse by 1 beam FWHM in each dimension until sufficient off-source pixels were present, or until the ellipse passed 10 FWHM in each dimension in which case the source was discarded as being part of an extended Stokes $I$ object.

Within the off-source pixels, the outlier-resistant\footnote{ Points more than twice the median absolute deviation away from the median were rejected before computing the mean and standard deviation. This is approximately equivalent to sigma clipping with a 3-sigma cutoff, but is more robust against extreme outliers which removes the need for iterating the clipping procedure.} mean and standard deviation were calculated for each channel's Stokes $Q$ and $U$. The means were taken as the foreground polarization around the source ($Q_{fg},U_{fg}$), and the root-mean-square of the standard deviations (averaging over channels and Stokes $Q$ and $U$) was taken as the local value for the noise combined with the uncertainty in the foreground estimate.

Within the on-source pixels, the noise ($\sigma_{QU}$) was calculated pixel-wise as the quadrature sum of the local noise determined from the off-source pixels and an instrumental leakage term, which was set equal to 0.3\% of the Stokes $I$ map \citep{Brown2002}:
\[\sigma_{QU} = \sqrt{\sigma_\mathrm{noise}^2 + (0.003 I)^2}. \] 
The foreground-subtracted and de-biased polarized intensity was calculated per channel, using a modified form of the  equation of \citet{Wardle1974}, as
\[P=\sqrt{(Q-Q_{fg})^2 + (U-U_{fg})^2 - \sigma_{QU}^2}.\]
This value was averaged over the 4 channels, and then divided by $\sigma_{QU}/\sqrt{4}$ (where the factor of $\sqrt{4}$ accounts for the decreased noise in the channel-averaged map) to produce a polarized intensity signal-to-noise ratio map. If any of the on-source pixels were found to have a polarized signal-to-noise ratio greater than 5, the source was classified as a candidate polarized source, and was processed through the RM determination and additional testing described in the next section.

This procedure produced a large number of false positives. In many cases, polarized pixels belonging to other neighboring sources would be included as on-source pixels, causing sources with polarized neighbors to be labeled as candidates. Another common cause was sources associated with extended Stokes $I$ emission, which would cause many pixels to be above the threshold and thus count as on-source, increasing the odds of encountering a high-polarization outlier in the noise distribution. Since polarized intensity does not follow Gaussian statistics, especially if interferometric image artifacts are present, the S:N>5 threshold is not as restrictive as might be ordinarily expected \citep{George2012}. We considered this high number of false positives as tolerable, since subsequent quality control tests done after RM determination were effective in rejecting these candidates.

\subsection{RM determination}
Candidate polarized sources were processed very similarly to the method of \citet{Brown2003}. To accurately constrain the on-source pixels, a 2D Gaussian was fit to the location of the source in the foreground-subtracted and debiased polarized intensity map calculated in the previous step. For a small number of candidates ($\sim$10\%) the Gaussian fitter failed to converge; these were visually inspected and found to be cases where the foreground subtraction was ineffective and had left significant diffuse polarization; these sources were immediately rejected. A few sources (<1\%) were identified as resolved with multiple spatially-separated polarized components; in these cases we fitted each component independently.

Since the Gaussian fitting routine was allowed to freely vary the centroid of the fit, it would occasionally lock onto a neighboring polarized source or additional polarized component within the same source, if it was stronger than the source that was the intended target of the fit. In cases where this occurred, the neighboring polarized emission was blanked out (by setting the polarized intensity to zero in the relevant pixels) and the fitting procedure repeated; this was iterated until the fit locked onto the correct source.

For the purposes of RM determination, pixels within the FWHM region of the fitted Gaussian, and having a polarized intensity signal-to-noise ratio greater than 5, were used. For each of these pixels, the foreground Stokes $Q$ and $U$ parameters were subtracted and the polarization angles of the remaining on-source polarization were calculated. These angles were subject to the `$n\pi$' ambiguity that occurs for Faraday-rotated polarization angles, where a change in the measured polarization angle by 180\degr\ would `wrap' back into the [0\degr,180\degr) range. Given our channel spacing, a change in polarization angle of $\pm90$\degr\ (half a wrap) would require an RM of $\pm$1840 \radu, which is much larger than we expect in this region of the Galactic disk. We assumed that the change in polarization angle between adjacent channels will always be less than 90\degr, and applied corrections of $\pm$180\degr\ as appropriate to the polarization angles to satisfy this condition.

Using the unwrapped polarization angles, a linear fit to polarization angle as a function of $\lambda^2$ was performed giving per pixel the fitted RM, error in RM, and probability-of-fit. The pixel-wise RM was averaged over the on-source pixels, weighted by the inverse square of the error in RM, to give the source RM. To assess the consistency of the RM across all the source pixels, a reduced $\chi^2$ value was calculated from the pixel-wise differences from the mean. The maximum (foreground subtracted, debiased) polarized intensity and Stokes $I$ from the on-source pixels were also determined, as well as the pixel-averaged fractional polarization.

After all of these values were computed, a series of quality control tests was applied to check for problems with the source. First, sources with fewer than 5 on-source pixels were rejected, where on-source was defined as being within the fitted FWHM and having polarized intensity signal-to-noise ratio greater than 5. Sources for which the fractional polarization was greater than 100\%, which could happen when the Gaussian fit locked onto a diffuse polarization feature near an unpolarized Stokes $I$ source were rejected. Sources with fractional polarization below 2\% were likewise rejected, to avoid any residual instrumental polarization leakage from being identified as a polarized source.

We observed that there was a population of partially resolved sources with strong RM gradients, for which a single value could not be assigned; the $\chi^2$ value calculated from the pixel RM averaging procedure was used to confirm that the source had a single well-defined RM and did not have any significant RM gradients across the on-source pixels. We defined a threshold for each source (dependent on the number of pixels, and thus the number of free parameters) corresponding to a 95\% confidence level in the $\chi^2$ distribution, meaning that a source with a single RM would fall below that threshold in 95\% of cases. Sources with $\chi^2$ values above this threshold were considered to have too much RM variation across the pixels and were rejected.

To test for Faraday complexity, we evaluated the linearity of the relationship between polarization angle and $\lambda^2$ using the probability-of-fit metric returned by the linear fitting routine. The probability-of-fit was averaged over the on-source pixels, and if the average probability was below 10\% the source was rejected for not having clear one-component Faraday-thin behavior.

An additional test for Faraday complexity was performed using the fractional polarization, which is expected to not vary with frequency for a Faraday-simple source \citep{LeRoux1961}. We determined the channel-averaged fractional polarization, then performed a $\chi^2$ test on the residuals (sum of the squares of the channel-wise differences from the mean). We rejected sources with $\chi^2$ values above the 95\% confidence level as being possibly Faraday-complex.

Sources that passed all of these criteria were then subjected to a manual inspection. This inspection verified a few conditions that were difficult to test in an automated way.  First, it was confirmed that the fitted FWHM region of the Gaussian fit was inside the boundaries of the Stokes $I$ source; sources for which a significant portion of the polarization was outside the Stokes $I$ counterpart were rejected. Second, it was manually verified that the source's polarized intensity was statistically significant, by confirming that the polarized intensity of the source stood out from the surrounding off-source pixels; sources where the on-source polarized intensity was indistinguishable from the surroundings were rejected.

The initial processing concluded with this step, but inspection of the resulting catalog showed that sources at the edges of the mosaics were less reliable, i.e. considerably more likely to have an RM significantly different from neighbouring sources. To maximize the reliability of our catalog, we removed the sources at the edges of the mosaics where the primary beam sensitivity was low, even if they passed all other tests. We used a threshold of 20\% of peak sensitivity, as given by the mosaic weight maps; sources below this threshold were discarded from the catalog.

At this stage all sources that were not rejected for one or more reason were considered as valid polarized sources with well-defined RMs. These sources went through a second manual inspection step, in which their polarized and Stokes $I$ morphology were inspected and recorded for source classification purposes. We identified each source as either resolved or unresolved in polarized intensity and Stokes $I$, as well as whether the source had any neighbouring sources within approximately 1\arcmin, and whether those neighbouring sources were also polarized. For resolved sources, we evaluated if the polarized intensity morphology matched the Stokes $I$ morphology. For unresolved sources, we also noted if there was an offset of at least 1 pixel (0.3\arcmin) between the polarized intensity peak and the Stokes $I$ peak, as this was seen in several sources. The source classifications were included in the final catalog.

\section{Final catalog}\label{sec:catalog}

Before assembling the final catalog, it was necessary to remove duplicate detections. Many sources were processed multiple times, due to the overlap between adjacent CGPS mosaics. Since the overlap regions were produced from the same observations, they were not independent measurements and it was not appropriate to combine the multiple detections together. In each case we chose to keep the detection that was farthest from the edge of the mosaic.

Finally, we checked our list of RMs against the known pulsars in this region of the sky, using the ATNF pulsar catalog \citep{Manchester2005}. Using a cross-matching radius of 0.9\arcmin\, we found 4 pulsars in our sample: B0355+54, J2007+2722, B2111+46, J2229+6114. 
For the remainder of this paper we assume that all the sources remaining in the main catalog are extragalactic.

The final catalog contains 2234 RMs (including the 4 pulsars), distributed over approximately 1300 square degrees. Selected columns of the catalog are shown in Table \ref{table:catalog_example}. The structure of this table is based on a in-development version of a new standardized format for reporting RMs (Van Eck et al, in prep).\footnote{Details of this format can be found at https://github.com/Cameron-Van-Eck/RMTable} This table reports the following quantities which vary by source:
\begin{itemize}
\item A source ID number, running from 1 to 2230 for the non-pulsar sources and giving the pulsar name for each of the 4 pulsars.
\item Positions in Galactic and equatorial coordinates. The positions are defined as the coordinates of the pixel closest to the Stokes $I$ source location, as reported by \citet{Taylor2017} or AEGEAN. Since these positions are quantized in units of the pixel size (0.005\degr), the uncertainty in position is half the pixel size.
\item Rotation measure and associated error, calculated as described above.
\item Polarized intensity, determined as the highest (foreground subtracted and debiased) polarized intensity of the on-source pixels
\item Stokes $I$ intensity, determined as the highest Stokes $I$ value of the on-source pixels.
\item The fractional polarization, determined as the average over the on-source pixels.
\item The beam major axis, which depends on declination, determined as 58\arcsec\ cosec $\delta$ (the beam minor axis and position angle were constant for all sources).
\item Source classification: the 4 pulsars are labelled as such, all other sources remain unclassified
\item Morphology flags from the manual inspection, defined as follows. These are stored as `Flag A value' in the table.
\begin{itemize}
\item C = Compact
\item R = Resolved (in Stokes $I$ and polarization, unless otherwise flagged)
\item S = Subset (polarized intensity morphology is smaller than the Stokes $I$ extent)
\item N = Neighbouring source within approximately 1\arcmin
\item P = Additional polarized component(s) seen in source or in neighbour
\item O = Offset (Polarized peak location does not match Stokes $I$ peak)
\end{itemize}
\item The CGPS mosaic in which the source was found, stored in the `Flag B value' column
\item The observation date, defined as the center of the approximately two-month period over which each mosaic was observed. This should be treated as approximate, as some mosaics had a few fields re-observed at a later time, and also the observation dates were supplied to us quantized to the nearest month.
\end{itemize}

Table \ref{tab:fixed_columns} reports values that apply to all rows in the catalog, which are included in the catalog for conformity with the standard format. Columns that are part of the standard that are not included in either of Tables~\ref{table:catalog_example} or \ref{tab:fixed_columns} are not supplied and have their default (blank) values as defined in the standard.

\begin{deluxetable*}{cccccccc}
\label{table:catalog_example}
\tablecaption{Source-Dependent Columns from the Catalog Table} \rotate
\tabletypesize{\footnotesize}
\startdata
\\
Catalog ID & RA & Dec & $l$ & $b$ & RM & RM Error & PI \\
& [\degr] & [\degr] & [\degr] &[\degr] & [rad m$^{-2}$] & [rad m$^{-2}$] & [mJy/beam]\\
\hline
1 & 23.46 & 63.02 & 127.72 & 0.54 & -103.5 & 14.9 & 7.87 \\
2 & 27.95 & 61.28 & 130.12 & -0.745 & -95.5 & 10.1 & 6.27 \\
3 & 24.82 & 60.93 & 128.71 & -1.4 & -185.7 & 12.2 & 8.33 \\
4 & 28.28 & 62.74 & 129.93 & 0.71 & -436.2 & 18.6 & 5.50 \\
5 & 21.75 & 60.19 & 127.34 & -2.375 & -68.4 & 17.1 & 6.04 \\
6 & 21.90 & 61.58 & 127.22 & -0.995 & -191.7 & 18.6 & 7.84 \\
... & ... & ... & ... & ... & ... & ... & ... \\
B0355+54 & 59.72 & 54.22 & 148.19 & 0.81 & 35.0 & 12.0 & 8.5 \\
J2007+2722 & 301.81 & 27.38 & 65.71 & -2.695 & -194.7 & 43.6 & 2.04 \\
B2111+46 & 318.35 & 46.74 & 89.005 & -1.265 & -138.8 & 24.6 & 3.29 \\
J2229+6114 & 337.25 & 61.24 & 106.64 & 2.96 & -204.2 & 6.6 & 9.3 \\ \hline \hline
Catalog ID  & Stokes $I$ & Fractional & Beam major & Source & Morphology & \multicolumn{2}{c}{Median observing}  \\
& [mJy/beam] &  polarization &axis [\degr] & type& &\multicolumn{2}{c}{epoch [MJD] } \\ 
\hline
1 & 86.70 & 0.089 & 0.0181 &  & C & \multicolumn{2}{c}{49991} \\
2 & 101.86 & 0.067 & 0.0184 &  & C & \multicolumn{2}{c}{49838} \\
3 & 194.25 & 0.049 & 0.0184 &  & C & \multicolumn{2}{c}{49838} \\
4 & 80.05 & 0.069 & 0.0181 &  & C & \multicolumn{2}{c}{51087} \\
5 & 141.93 & 0.045 & 0.0186 &  & C & \multicolumn{2}{c}{49991} \\
6 & 245.04 & 0.034 & 0.0183 &  & C & \multicolumn{2}{c}{49899} \\
... & ... & ... & ... & ... & ... & \multicolumn{2}{c}{... } \\
B0355+54 & 20.78 & 0.415 & 0.0199 & Pulsar & N & \multicolumn{2}{c}{51726} \\
J2007+2722 & 2.85 & 0.722 & 0.0350 & Pulsar & C & \multicolumn{2}{c}{53065}\\
B2111+46 & 22.34 & 0.151 & 0.0221 & Pulsar & N & \multicolumn{2}{c}{50722} \\
J2229+6114 & 23.41 & 0.365 & 0.0184 & Pulsar & RSP & \multicolumn{2}{c}{51057}\\
\enddata
\tablecomments{The full catalog is available in FITS format through the arXiv as ancillary data, and will be also online at the {\it Astrophysical Journal Supplements} and {\it Vizier} after publication. The structure of this table is based on the proposed standardized format for reporting RMs. The table shown here is just a portion for guidance regarding its form and content. In addition to the source-dependent values shown here, the .fits version has the source-independent values given in Table 3.}
\end{deluxetable*}

\begin{table*}
    \centering
    \begin{tabular}{|r|l|} \hline
        \bf{Column} & \bf{Value} \\ \hline
        RM determination method & `EVPA-linear fit'\\
        Ionospheric correction method & `None'\\
        Stokes $I$ reference frequency & 1.4207809 GHz\\
        Polarization reference frequency & 1.4207809 GHz\\
        Beam minor axis & 0.013611\degr\ (58\arcsec)\\
        Beam position angle & 0\degr\\
        Polarization bias correction method & `1974ApJ...194..249W' 
        \\ \hphantom{ } & \citep{Wardle1974}\\
        Peak or integrated flux? & `Peak'\\
        Minimum frequency & 1.407194 GHz\\
        Maximum frequency & 1.43463 GHz\\
        Channel width & 7.5 MHz\\
        Number of channels & 4\\
        Telescope & `DRAO-ST'\\
        Interval of observation\tablenotemark{a} & 60 days\\
        Catalog & `New CGPS (Van Eck et al 2020 in prep)'\tablenotemark{b}\\
        Flag A name & `Morphology'\\
        Flag B name & `Mosaic name'\\
        \hline
    \end{tabular}
    \tablenotetext{a}{Defined as the interval between the first and last observations used. This value is approximate; most fields were observed repeatedly over a period of two months with different baseline configurations to produce the final images.}
    \tablenotetext{b}{This will be changed to the bibcode of this paper after publication.}
    \caption{Source-independent Columns in the Catalog Table\label{tab:fixed_columns}}
\end{table*}

\begin{figure*}[p]
    \centering
    \includegraphics[width=\linewidth,height=0.8\textheight,keepaspectratio]{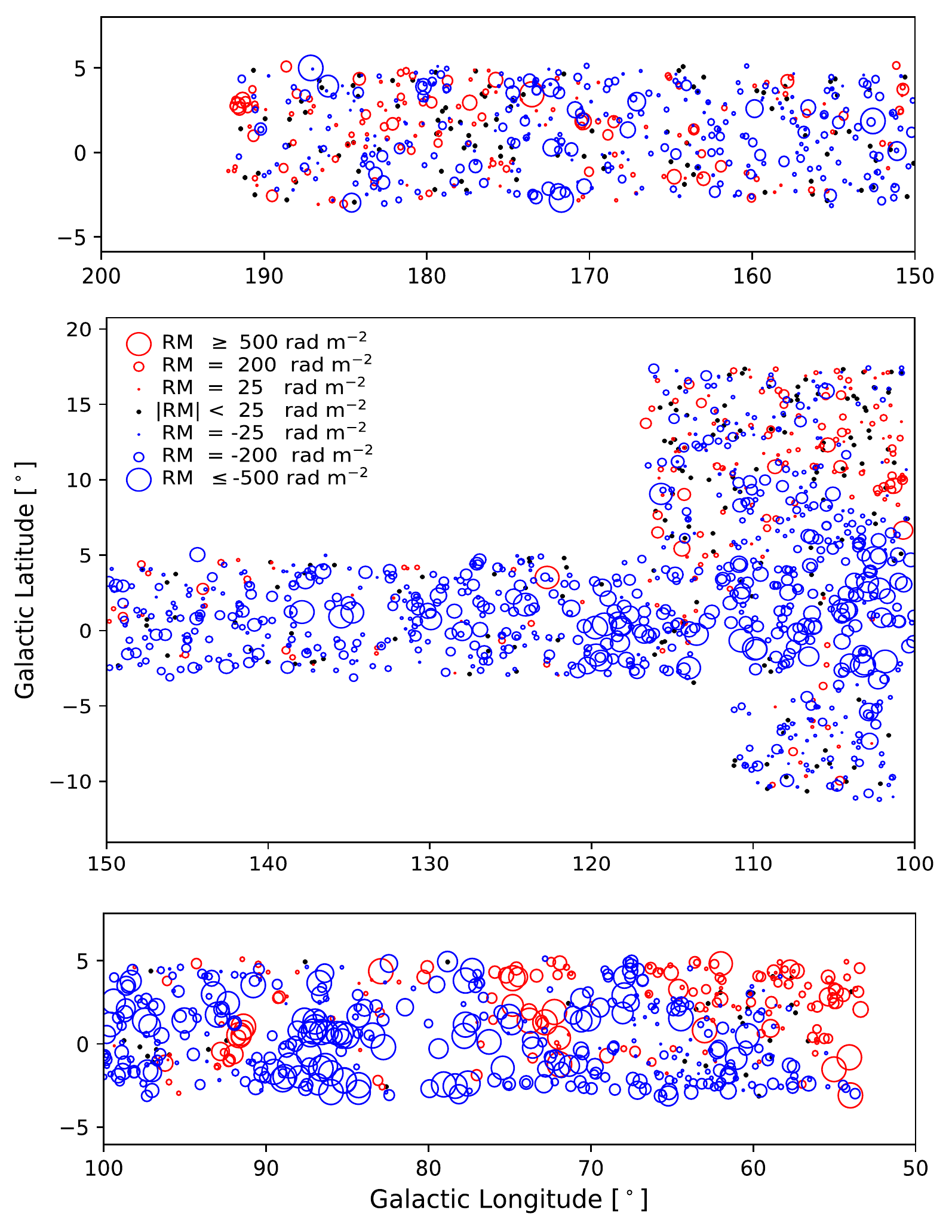}
    \caption{The locations and RMs of polarized sources found in the CGPS region. Circle diameter is proportional to the magnitude of the RM (capped at 500 \radu), with positive and negative RMs colored in red and blue respectively. RMs with magnitude smaller than 25 \radu\ are dispayed as a filled black circle.}
    \label{fig:circleplot}
\end{figure*}

Figure \ref{fig:circleplot} shows the positions and RMs of all the non-pulsar sources. The density of sources is relatively uniform with two exceptions: a region around $l$=80\degr\ where the bright Cygnus X region dominates Stokes $I$ and impacted source finding, and a smaller region around $l$=112\degr, $b$=-2\degr\ where the bright supernova remnant Cas A significantly increases the noise levels.

\subsection{Comparison with previous catalogs}
\begin{figure}[htb]
    \centering
    \includegraphics[width=\linewidth,height=0.9\textheight,keepaspectratio]{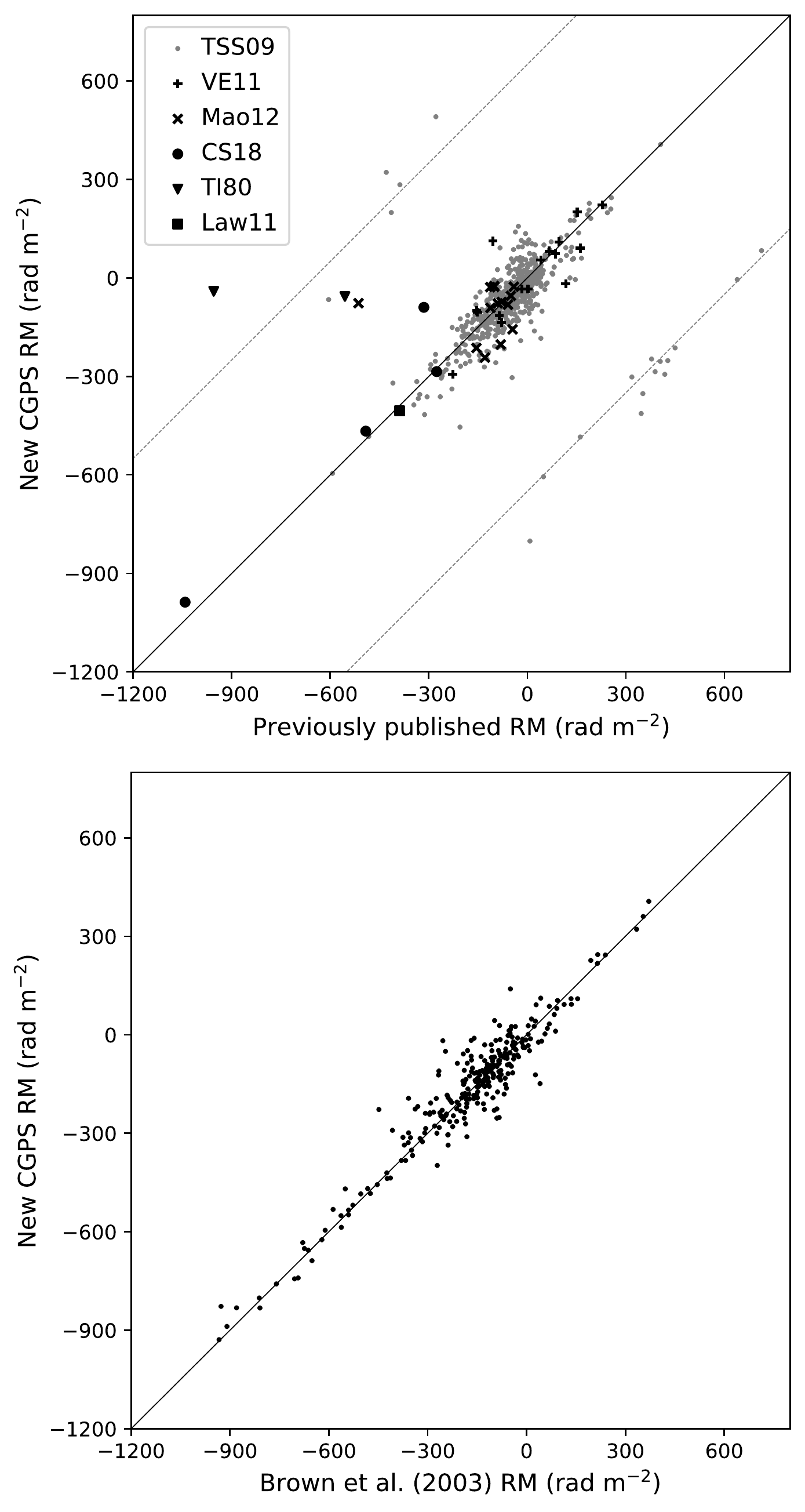}
    \caption{{\em Top: }Comparison of reported RMs for sources present in our catalog and previously published catalogs. The solid line marks 1:1 agreement between the catalogs, while dashed lines show a $\pm$650 \radu\ offset corresponding to the $n\pi$ ambiguity in the \citet{Taylor09} catalog. The catalog abbreviations are given in the text. {\em Bottom:} As the top plot, but only comparing against the \citet{Brown2003} RM catalog.}
    \label{fig:comparison}
\end{figure}

To assess the quality of our catalog, we compared the RMs against a new master catalog of previously published RMs.\footnote{This master catalog is being assembled as part of an effort to standardize the reporting of RMs between different projects (Van Eck et al, in prep). We used version 0.1.8 of this catalog, which can be found at https://github.com/Cameron-Van-Eck/RMTable.} The previously published CGPS RMs \citep{Brown2003} were deliberately excluded and will be considered separately.
Using a cross-matching radius of 30\arcsec, 601 matches were found: 564 from the catalog of \citet[TSS09 in Fig \ref{fig:comparison}]{Taylor09}, 16 from \citet[VE11 in Fig \ref{fig:comparison}]{VanEck11}, 14 from \citet[Mao12 in Fig \ref{fig:comparison}]{Mao2012}, 4 from \citet[CS18 in Fig \ref{fig:comparison}]{Costa2018}, 2 from \citet[TI80 in Fig \ref{fig:comparison}]{Tabara1980}, and 1 from \citet[Law11 in Fig \ref{fig:comparison}]{Law2011}. The comparison of their RM values to ours is shown in the top panel of Figure~\ref{fig:comparison}.  No sources were found that had counterparts within two or more previous catalogs.

In general there is qualitative agreement between the new RMs and previous measurements. There is a small population of sources from \citet{Taylor09} that are offset by approximately $\pm 650$ \radu\ because of a known problem with the angle unwrapping ambiguity in their algorithm. These sources appear near the two dotted lines in the top panel of Fig.~\ref{fig:comparison}.

A few sources show conspicuously large deviations between new and old RM measurements. Both of the sources that were also present in the \citet{Tabara1980} catalog have very different RMs (>100 \radu\ difference), as do two of the sources from \citet{Costa2018}. However, all 4 of these sources show signs of Faraday complexity in the older measurements: significant changes in polarized fraction at different frequencies for the \citet{Tabara1980} RMs, and large linear fit residuals for the \citet{Costa2018} RMs. This can explain how different RMs can be observed over different frequency ranges. We interpret this as further evidence that individual source RMs should be interpreted carefully whenever the presence of Faraday complexity cannot be reliably ruled out. In addition to these sources, one of the \citet{Mao2012} RMs is also significantly different from ours, but no reason for this difference can be easily identified.

We further investigated the differences between our catalog and the matching sources in the \citet{Taylor09} catalog to search for possible systematic effects. When testing for correlations between the absolute value of the RM difference (between the two catalogs) and source properties, we found a significant anti-correlation with the signal-to-noise ratio (Spearman $\rho$ = -0.36, p$\approx10^{-19}$), as well as with related quantities such as polarized intensity. However, this correlation was no longer present ($\rho$ = -0.027, p=0.51) after the differences were normalized by the uncertainty in the difference (the quadrature sum of the uncertainties in the individual RMs). We interpret this to indicate that the RM uncertainties in both catalogs have the correct dependence on signal-to-noise ratio. We find no significant correlations with the uncertainty-normalized RM differences.

\subsection{Comparison with the 2003 CGPS RM catalog}
In addition to comparing against other observations, we also compared this new CGPS catalog against the initial CGPS catalog \citep{Brown2003} to determine how significantly the change in processing affected the resulting RMs. Of the 380 sources in the 2003 catalog, all were present in the \citet{Taylor2017} catalog, and all but four were identified as polarized source candidates. Those 4 were found to be located in regions of strong, position-dependent diffuse polarized emission. This caused our foreground subtraction algorithm to assign them high uncertainties. Thus, even though clear on-source polarization could be seen by a visual inspection, our algorithm classified them as below the signal to noise threshold and they were discarded.

Of the remaining 376 sources, 95 did not pass the quality control tests of the new pipeline. Four were found to fail Gaussian fitting in polarized intensity (probably as a result of the improved foreground subtraction), 15 were found to not have enough pixels above the signal-to-noise threshold (as a result of the new noise calculations), 11 failed the pixel-averaging $\chi^2$ test and 15 failed the linearity probability-of-fit test (for both tests, as a result of the new noise and error estimates). These sources, and their reported results from both the new and original pipelines, were inspected, and we found that nearly all of these were close to the pass/fail thresholds in the original pipeline and were pushed across one or more of these thresholds by the changes to the noise and foreground calculations. The $\chi^2$ test for constant fractional polarization was not used in the previous catalog; this test caused 43 previous sources to fail. Seven sources were found to be outside of the 20\% sensitivity threshold we used to remove edge sources with lower reliability.

The remaining 281 passed all tests in the improved pipeline and are included in the new catalog with updated RMs; a comparison of the updated RMs to those of the original appears in the lower panel of Fig.~\ref{fig:comparison}. We strongly caution users of this catalog that these RMs were derived from the same observations as the 2003 catalog; they are not independent measurements and should not be combined with the previous catalog for statistical analyses. We recommend that our catalog completely replace the original CGPS catalog for all future analyses.

\section{Analysis}\label{sec:analysis}

In this section we look at a few examples of analyses that can be done with the new RMs, including a short description of some work that has already been published using preliminary versions of these data.

\subsection{Large-scale trends}\label{sec:trends}
In Figure \ref{fig:longitude} we show the statistical properties of the rotation measures as a function of longitude. The mean RM ($\overline{\text{RM}}$, top panel) generally shows smooth trends with longitude: the outer Galaxy region 130\degr\ < $\ell$ < 180\degr\ has a smooth and steady trend in RM from negative towards zero with increasing longitude. This indicates that the large-scale magnetic field is embedded in a ubiquitous phase of the ISM. This was confirmed by \citet{Foster13}, who used a preliminary version of this catalog and found a strong relation between RMs of extragalactic sources and the optically thin hydrogen column density in the same direction, indicating that the warm neutral medium is the main carrier of the large-scale magnetic field in the Galaxy. 
RMs cross zero near the anti-center, indicating that the magnetic field is nearly azimuthal in the outer Galaxy \citep{VanEck11}. This is supported by the study of the supernova remnant G182.4+4.3, close to the anti-center, in which the ambient field is almost perpendicular to the line of sight \citep{Kothes09}. In the inner Galaxy the magnetic field closely follows the spiral arms, and, starting near $\ell$< 70\degr\, the RMs show a strong swing from negative to positive with decreasing longitude. This is the reversal of the field between the local and Sagittarius arms \citep{VanEck11}.

The third panel of Figure~\ref{fig:longitude} shows the longitude-dependence of the ratio between the standard deviation ($\sigma_{\text{RM}}$) and the absolute value of the mean of the RMs within the 2 degree bins we used to compute the statistics. We note that over a substantial extent of the longitude range, from $l\approx$ 100\degr\ to $l\approx$ 150\degr, the ratio $\sigma_{\text{RM}}/|\overline{\text{RM}}|$ is well constrained, with a correlation coefficient of 0.57 (p=0.003) between $\sigma_{\text{RM}}$ and $|\overline{\text{RM}}|$. This is consistent with the similar result presented in \cite{Brown2001} using RM data from the first phase of the CGPS \citep{Taylor2003}. This correlation was interpreted by \cite{Brown2001} to indicate a preferential alignment between the small- and large-scale magnetic field components, a precursor to the concept of an ordered-random GMF component \citep[e.g.,][]{Jaffe2010}. At longitudes where the RMs approach zero (near the anti-center and the large-scale reversal region), the ratios $\sigma_{\text{RM}}/|\overline{\text{RM}}|$ increase significantly as a consequence of dividing by very small values of $|\overline{\text{RM}}|$. Enhanced variability in the Cygnus X region (see \S \ref{sec:anomalies}) also contributes deviations from the near-constant ratio seen in the mid-longitude range. The mid-longitude range of the CGPS, where the magnetic field has a sufficient line-of-sight component to yield substantial RM values, is an ideal testbed for statistical studies of the connection between small- and large-scale GMF components. 

The variation of RM with Galactic latitude is shown in Figure~\ref{fig:latitude} for the two latitude extensions. The absolute value of RM falls significantly away from the mid-plane, and drops almost to zero at high latitudes, as expected. However, RM variation is not symmetric around the mid-plane. This asymmetry appears to be caused by the anomaly known as Region A {\citep{sk80}}, which is primarily affecting RMs south of the Galactic disk. A more thorough analysis of the latitude extensions, using an earlier version of the catalog, was done by \citet{Cooper2014}.

\subsection{Localized anomalies}\label{sec:anomalies}

\begin{figure}
    \centering
    \includegraphics[width=\linewidth,height=0.9\textheight,keepaspectratio]{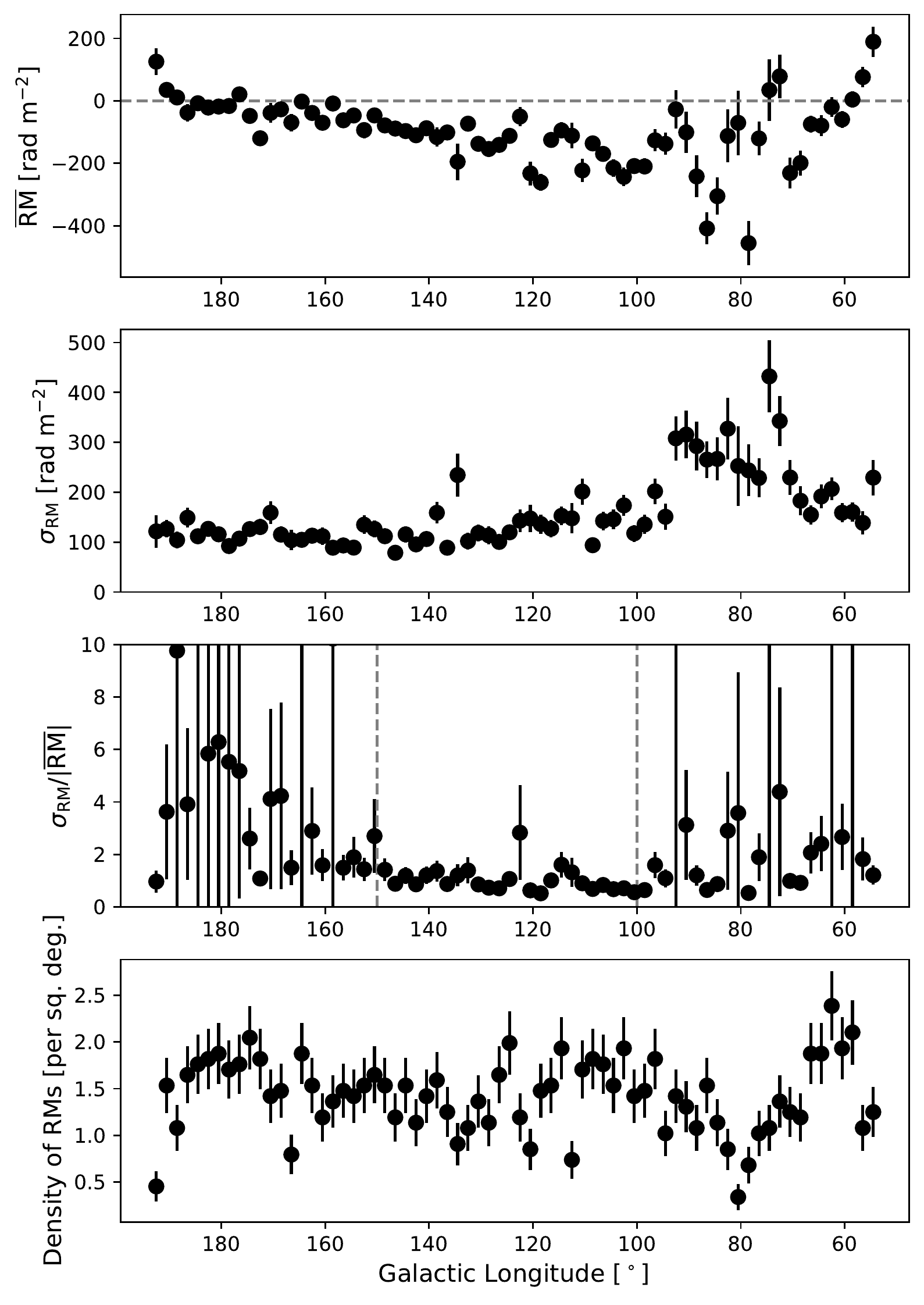}  
    \caption{Trends in the RM statistics as a function of longitude (computed over 2 degree bins), in the disk ($|b|<5.5\degr$). {\it Top:} mean of rotation measure; {\it second:} scatter (standard deviation) of rotation measure; {\it{third:}} ratio of the top two panels; {\it bottom:} source density of RMs. The error bars for the mean RM and the standard deviation were determined as the standard errors of the mean and standard deviation respectively. The error bars on the ratios take both of these into account using error propagation. The error bars on the source density were calculated assuming a Poisson distribution in the source counts.
    }
    \label{fig:longitude}
\end{figure}

\begin{figure}
    \centering
    \includegraphics[width=\linewidth,height=0.6\textheight,keepaspectratio]{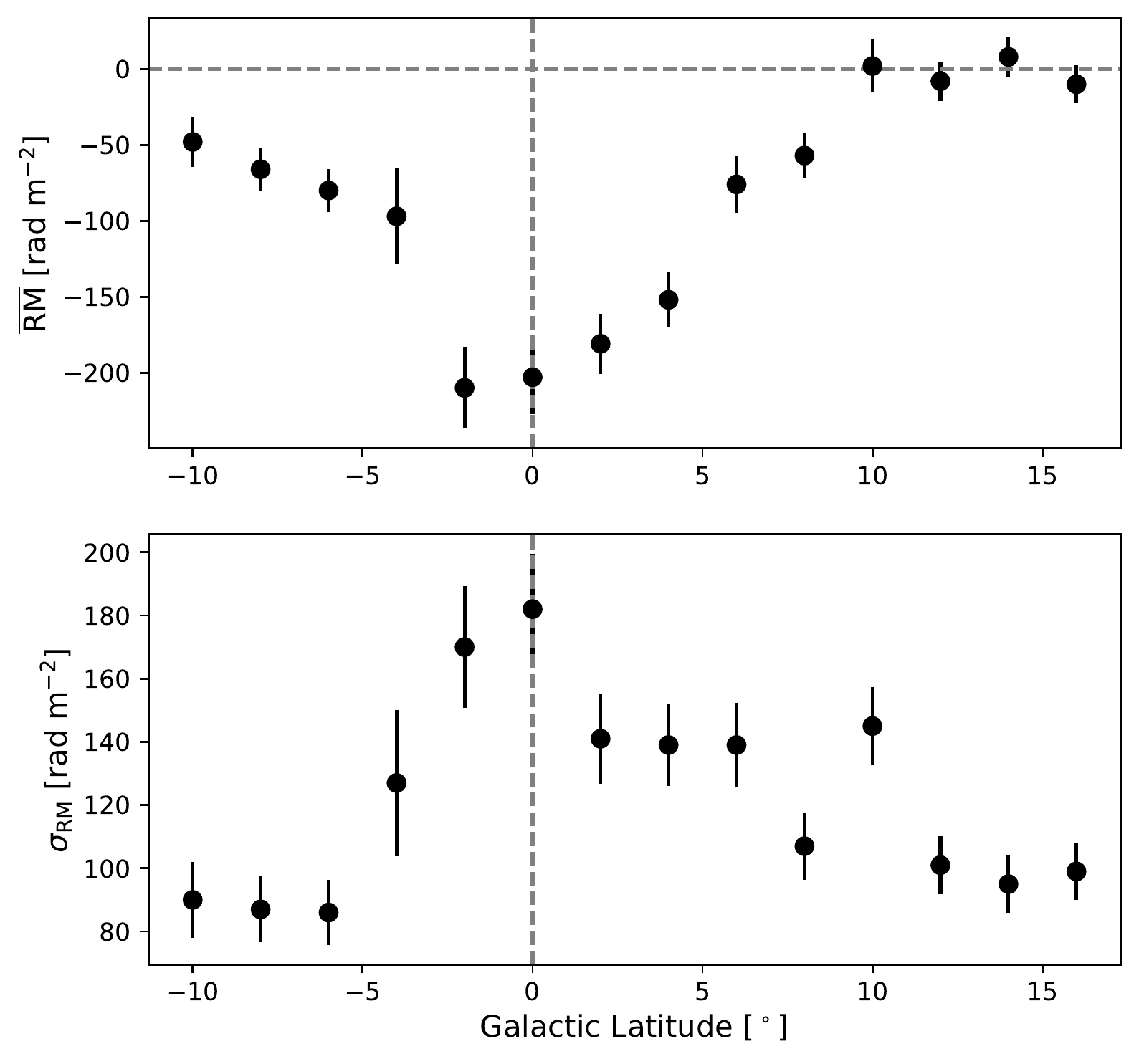} 
    \caption{Trends in the RM statistics as a function of latitude (computed over 2 degree bins), in latitude extension regions ($100\degr<l<117\degr$). {\it Top:} mean of rotation measure; {\it bottom:} scatter (standard deviation) of rotation measure. The Galactic mid-plane has been marked by vertical dashed lines.}
    \label{fig:latitude}
\end{figure}

\begin{figure*}
    \centering
    \includegraphics[width=\linewidth,height=0.9\textheight,keepaspectratio]{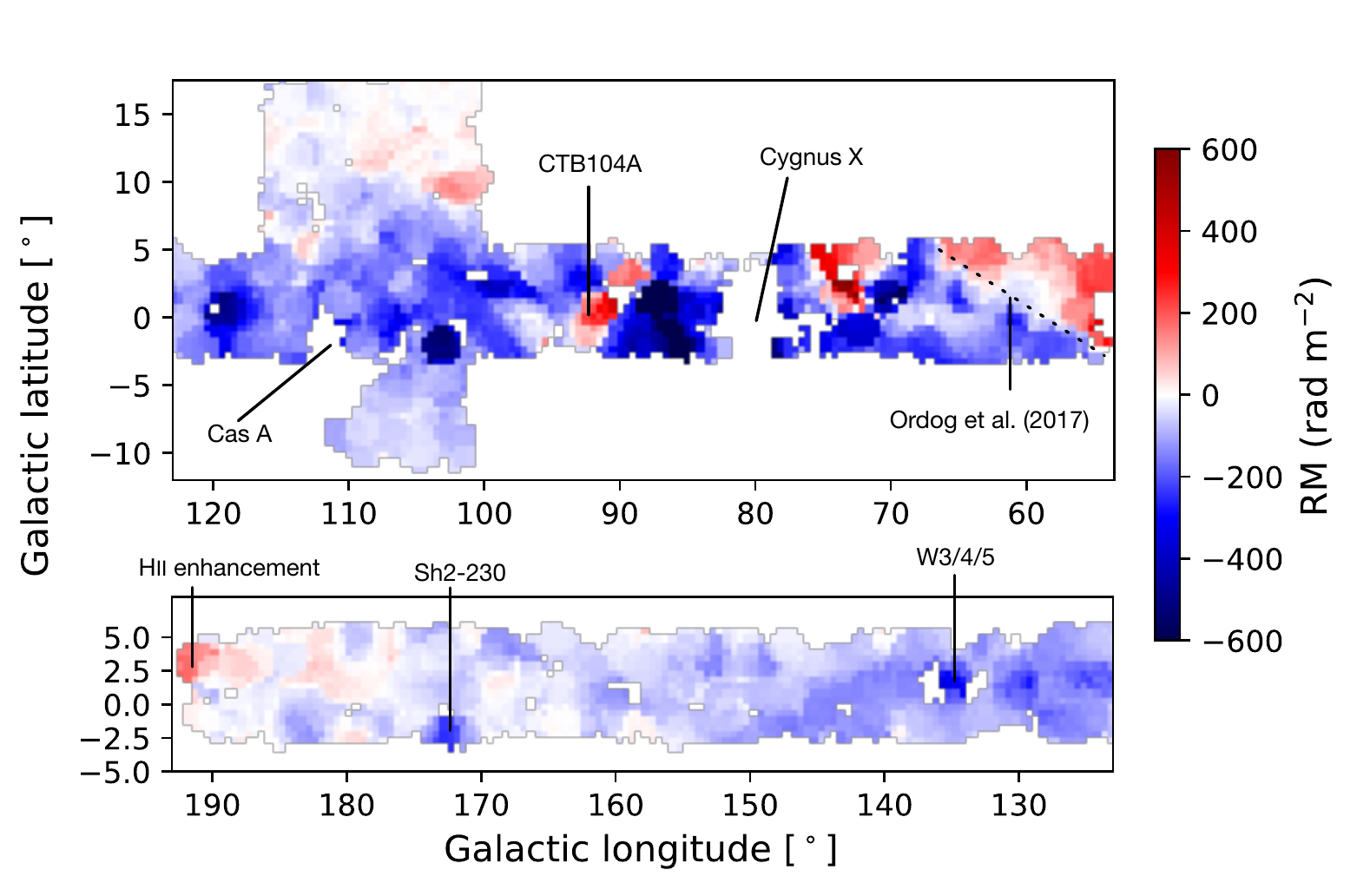}  
    \caption{Smoothed distribution of RM, computed per-position as the weighted mean of RMs within 2 degrees. Positions with less than 5 sources within 2 degrees are not shown. Individual regions described in the text are marked.}
    \label{fig:smoothed}
\end{figure*}

Several regions or individual points that deviate from the large-scale trends are also visible, and are usually reflected by larger variations in the RM within each bin (as shown by the second panel of Fig.~\ref{fig:longitude}). Most of these deviations are caused by smaller scale structures in the ISM, such as \ion{H}{2} regions. Several examples can be easily identified: the enhanced RM in the highest longitude bin (191.5--193.5\degr) is due to a cluster of sources behind an \ion{H}{2} enhancement at $l\approx$ 191.5\degr, $b\approx$ +2.8\degr\ with much larger positive RMs than the surrounding sources; the anomalously low mean RM, large scatter, and lower source density around $l$ = 172\degr\ is associated with a large \ion{H}{2} region complex centred on Sh2-230; the similar anomaly around $l$ = 135\degr\ is due to the W3/4/5 complex. Figure~\ref{fig:smoothed} shows the smoothed distribution of RMs with these regions highlighted.

Figure~\ref{fig:longitude} reveals a large scatter in RMs towards an area 15\degr\ wide centered on $\ell$ = 80\degr. This is the Cygnus X region where we are looking along our own spiral arm, the Orion spur. In Cygnus X we are looking through several layers of star forming regions and \ion{H}{2} regions \citep{Gottschalk2012}, and the heavy concentrations of ionized material explain the scatter in RM. RM source density is low in this area (see Figure~\ref{fig:longitude}) because of the very strong extended emission, and the area is left blank in Figure~\ref{fig:smoothed}. Pulsar RMs towards Cygnus X are almost all positive above latitude $-$4 deg, at least in the range 75\degr\ < $\ell$ <82\degr\ \citep[Figure 9 of][]{Kothes2020}. The region of positive RMs around $\ell$ = 75\degr, {$b$ = +4\degr} may be associated with the low-longitude side of Cygnus X or may be associated with the positive RMs found at lower longitudes. A study of compact source RMs and a comparison with the RM of extended emission (as in \citealt{Ordog2019}) would contribute to understanding this region.

Other departures from the overall trend can be seen in Figures~\ref{fig:circleplot} and \ref{fig:smoothed}, but are less obvious in Figure ~\ref{fig:longitude}. We can identify these features with \ion{H}{2} regions or supernova remnants through comparison with CGPS images of the total-intensity along the Galactic plane; we use the 408-MHz maps of \citet{Tung17}, especially their Figures 6 to 12:
\begin{itemize}
\item A region of strong negative RM near $\ell$ = 143\degr, 0\degr\ < $b$ < 2\degr\ is associated with the W3/4/5 \ion{H}{2} region complex.
\item A region of strong negative RM near $\ell$ = 172\degr, $b$ = $-$3\degr\ is associated with the \ion{H}{2} complex Sh 2-230. This complex spans almost the whole latitude range of the survey, accounting for a low density of RMs in this area.
\item At $\ell$ = 93\degr, $b$ = 0\degr\ we see an area of positive RM where the surrounding RMs are negative. This feature was reported by \citet{Clegg1992} and was noted in the 2003 CGPS RM catalog \citep{Brown2003}. This anomaly is related to the supernova remnant CTB104A and/or its environment \citep{Uyaniker02}.
\end{itemize}

There is also an area with very low RM magnitudes embedded within large RMs slightly below the mid-plane at approximately $62\degr < {\ell} < 70\degr$ (see Figure  \ref{fig:circleplot}). The reason for this is unclear. This region is largely empty in Stokes I, from the edge of Cygnus~X to approximately where Sagittarius arm begins, with no obvious Stokes I counterpart to these low RMs.


{}

{}

\subsection{Comparison with diffuse emission}
One weakness in a latitude-averaged analysis like Fig.~\ref{fig:longitude} is that transitions that are not parallel to the Galactic plane can be lost or mis-represented. \citet{Ordog2017} used an earlier version of this catalog to identify that the transition in the sign of RM around $l$ = 60\degr\ is in fact not aligned with the Galactic plane but is clearly along a diagonal line (their Figure 2, reproduced in Figure~\ref{fig:smoothed}). They also found that this transition was reflected in the diffuse polarized emission present in the CGPS data. 

\citet{Ordog2019} used a pre-final version of this RM catalog (which was made without the 20\% sensitivity threshold and fractional polarization variation tests) to compare with rotation measures derived from the diffuse polarized emission that is also present in the CGPS data. They found a strong correspondence between the diffuse emission RMs and the extragalactic source RMs, except in a few regions (most of the regions that are discussed in Sects.~\ref{sec:trends} and \ref{sec:anomalies}) where smaller local features strongly influence the RMs. They found that through most of the lines of sight the extragalactic RMs (which probe the full line of sight through the Milky Way) were approximately twice as large as the RMs of the diffuse emission (which is distributed through the Milky Way in a complex way). The details of this result and their interpretation are not repeated here and can be found in their paper.

\citet{Stutz14} performed a power spectrum analysis of the CGPS diffuse polarized emission at a resolution of {2.67\degr}. In the high-latitude extension, they found a sharp transition to a steeper power law index at a latitude of about {+9\degr}, which indicates a transition from small-scale to large-scale structures. This is interpreted as the disk-halo interface and agrees with the location where the RMs reduce to approximately zero (Figure~\ref{fig:latitude}).


\section{Summary and Conclusions}\label{sec:summary}
The Canadian Galactic Plane Survey (CGPS) covers a large area of the Galactic plane, from $l$=52\degr\ to $l$=192\degr, and provides full polarization data in 4 closely spaced frequency channels, enabling measurements of Faraday rotation of background sources. We have expanded on the work by \citet{Brown2003} determining rotation measures from the CGPS data, with an improved processing pipeline that identifies more polarized sources and improves the foreground subtraction and error analysis. Applying this pipeline to the full CGPS region, including north and south latitude extensions, we have produced a new catalog of 2234 RMs covering approximately 1300 square degrees. Of these, 4 were identified as known pulsars. We have compiled this catalog following a forthcoming standard format to try to maximize the future value of the RMs. 

As a verification of the RM values, we identified 564 sources in our catalog with previously observed (independent) RMs, and found good agreement between our values and the previous measurements, with exceptions mostly identified as problems in the previous observations or as sources that showed signs of Faraday complex behaviour. We also compared the RMs we obtained against those from the original CGPS RM catalog \citep{Brown2003}, and found generally good agreement there as well. Of the original 380 sources, 95 were found to no longer pass the quality control tests in the new pipeline; most of these were cases where a source transitioned from marginally passing a test to marginally failing or as the result of failing the fractional polarization variation test that was implemented in the new pipeline. In addition, several sources present in both catalogs were found to have different RMs of order several tens of \radu. Since the underlying data for both sets of RMs are the same, we interpreted both the change in RM and the change in the quality control test outcomes to be due to differences in the foreground subtraction. While we are confident that our improved pipeline performs foreground subtraction as effectively as possible, we caution users that individual RMs may be subject to unquantified systematic errors of the order of a few tens of \radu. We also warn users that these RMs are not independent measurements from the \citet{Brown2003} catalog. We recommend that this new catalog replace the old one in future analysis.

With a typical source density of about 2 RMs per square degree, our catalog significantly improves on the two previous large surveys of this area, \citet{Brown2003} and \citet{Taylor09}, which have source densities of approximately 1 RM per square degree.  Our catalog is now the highest density large RM survey of the Galactic plane, and will be very useful for future studies of magnetic fields inside the Galactic disk. This includes both studies of large scale structure in the Galactic magnetic field \citep{Jaffe2019} as well as studies of magnetism in smaller scale objects \citep[e.g.,][]{Tahani2018}. Our catalog has already contributed significantly to several such studies \citep{Ordog2019,Ma2020}.

The next generation of rotation measure surveys is already underway. While POSSUM \citep{Gaensler2010} will not have the declination coverage to overlap with the CGPS region, VLASS \citep{Mao2014} will cover the CGPS region at higher frequencies. While the VLASS RM catalog is expected to have a higher source density than the CGPS, our catalog will be complementary with its coverage of lower frequencies. This will be useful as a probe of Faraday complexity, for example by searching for frequency dependence in the RM. We expect that this catalog will be a unique and useful resource for many years.

\acknowledgments
The authors would like to thank Diane Parchomchuk, Jack Dawson, and Ev Sheehan, who operated the telescope and prepared the data for analysis through the long period of the CGPS observations. We are indebted to them for their skill, and for their thorough and painstaking work.

This work has made use of the following software packages: NumPy \citep{Numpy}, Matplotlib \citep{Matplotlib}, Astropy \citep{astropy1,astropy2}, and the Karma visualization tools \citep{Karma}.

The Canadian Galactic Plane Survey is a Canadian project with international partners. The Dominion Radio Astrophysical Observatory is operated as a national facility by the National Research Council Canada. This research has been supported by grants from the Natural Sciences and Engineering Research Council.

The Dunlap Institute is funded through an endowment established by the David Dunlap family and the University of Toronto.

\bibliographystyle{apj}
\bibliography{References}

\end{document}